\documentclass[]{aastex631}

\usepackage{graphicx}% Include figure files
\usepackage{dcolumn}% Align table columns on decimal point
\usepackage{bm}% bold math
\usepackage{hyperref}% add hypertext capabilities
\usepackage{color}
\usepackage{fontawesome} % For github symbols etc
\usepackage{comment}
\usepackage{slashed}
\usepackage{tikz}
\usetikzlibrary{positioning,decorations.pathmorphing}
\usepackage{braket}
\usepackage{xspace}
\usepackage{amsmath}
\usepackage{amssymb}

\shorttitle{Dark extended lenses with Roman}
\shortauthors{DeRocco et al.}

\begin{document}

% \preprint{KEK-QUP-2023-0037}
% \preprint{KEK-TH-2586}
% \preprint{KEK-Cosmo-0335} 
% \preprint{IPMU23-0050} 

\newcommand{\be}{\begin{equation}}
\newcommand{\ee}{\end{equation}}
\newcommand{\rs}{R_s}
\newcommand{\rse}{r_s}
\newcommand{\rn}{R_{90}}
\newcommand{\rne}{r_{90}}

\definecolor{deepblue}{rgb}{0.2,0.2,0.8}
\definecolor{deepred}{rgb}{0.8,0.2,0.2}
\definecolor{deeporange}{rgb}{0.8,0.5,0.2}
\definecolor{deepgreen}{rgb}{0.2,0.8,0.2}
\newcommand{\red}[1]{\textcolor{deepred}{#1}}

\title{New Light on Dark Extended Lenses with the Roman Space Telescope}

\author{William DeRocco}
%\email{wderocco@ucsc.edu}
\affiliation{Santa Cruz Institute for Particle Physics, University of California, Santa Cruz, CA 95062, USA}

\author{Nolan Smyth}
%\email{nwsmyth@ucsc.edu}
\affiliation{Santa Cruz Institute for Particle Physics, University of California, Santa Cruz, CA 95062, USA}

\author{Volodymyr Takhistov}
%\email{vtakhist@post.kek.jp}
\affiliation{Theory Center, Institute of Particle and Nuclear Studies, High Energy Accelerator Research Organization (KEK), Tsukuba 305-0801, Japan}
\affiliation{International Center for Quantum-field Measurement Systems for Studies of the Universe
and Particles (QUP), KEK, 1-1 Oho, Tsukuba, Ibaraki 305-0801, Japan
}
\affiliation{Graduate University for Advanced Studies (SOKENDAI),
1-1 Oho, Tsukuba, Ibaraki 305-0801, Japan
}
\affiliation{Kavli Institute for the Physics and Mathematics of the Universe (WPI), Chiba 277-8583, Japan}

\correspondingauthor{William DeRocco}
\email{wderocco@ucsc.edu}

% \date{\today}

\begin{abstract}
The Roman Space Telescope's Galactic Bulge Time Domain Survey will constitute the most sensitive microlensing survey of the Galactic Bulge to date, opening up new opportunities to search for dark matter (DM). Many extensions of the Standard Model predict the formation of \textit{extended} DM substructures, such as DM subhalos, boson/axion stars, and halo-dressed primordial black holes. We demonstrate that for such targets, Roman will be sensitive to a broad parameter space up to four orders of magnitude below existing constraints. Our analysis can be readily applied to other extended DM configurations as well. 

\end{abstract}

% \maketitle

\section{Introduction}
\label{sec:intro}

Dark matter (DM) is the predominant constituent of all matter in the Universe. However, it has been detected only through its gravitational interactions and its nature remains mysterious~(e.g.~\citet{Gelmini:2015zpa,Bertone:2016nfn}). Significant efforts have been devoted to search for both particle DM~(e.g.~\citet{Bertone:2004pz}) as well as \textit{macroscopic} DM consisting of localized point-like objects, the canonical example being primordial black holes (PBHs) (see e.g.~\citet{Sasaki:2018dmp,Green:2020jor,Carr:2020gox,Escriva:2022duf} for a review). In the broad mass range of $\sim 10^{-10} - 10 \,M_{\odot}$, microlensing surveys by Subaru Hyper-Suprime Cam~\citep{Niikura:2017zjd}, OGLE~\citep{Niikura:2019kqi}, and MACHO/EROS~\citep{MACHO:1998qtf} have dominated the sensitivity reach for probing macroscopic DM.

Beyond point-like macroscopic DM, many motivated theories instead predict DM to be composed of \textit{extended} macroscopic configurations, which have received less attention. 
Examples include DM halo substructures~\citep{Erickcek:2011us,Barenboim:2013gya,Fan:2014zua,Dror:2017gjq} and axion miniclusters with axion stars~\citep{Kolb:1993zz, Eggemeier:2019jsu}.
Gravitational lensing has been identified as a promising method for probing extended DM substructures~\citep{Zackrisson:2009rc}, with recent analyses demonstrating that existing microlensing surveys can sensitively constrain a variety of models~\citep{Fairbairn:2017sil,Croon:2020ouk,Croon:2020wpr,Cai:2022kbp}, and probes complementary parameter space to other techniques, such as halometry from  astrometry \citep{2018JCAP...07..041V}.

The launch of the Nancy Grace Roman Space Telescope (Roman)~\citep{Spergel:2015} in 2027 will usher in a new era of astronomy. Among its high-priority science requirements is the Galactic Bulge Time Domain Survey (GBTDS), a microlensing survey with unprecedented sensitivity to non-luminous astrophysical bodies. Though this survey is nominally intended to provide a census of Galactic exoplanets, it also has the potential to provide a new window into DM at macroscopic scales. While its sensitivity to point-like DM sources has been discussed previously in the literature \citep{pruett_primordial_2022,DeRocco:2023gde,fardeen2023astrometric}, no prior study has addressed its sensitivity to extended DM distributions.

In this work, we perform the first sensitivity analysis of the Roman Space Telescope to extended DM structures. As we demonstrate,  
the GBTDS will be able to probe scenarios with such sub-structures many orders of magnitude below existing constraints, reaching fractional contributions to the total DM energy density as low as $f_{\rm DM} \sim 10^{-6}$ and exploring new, motivated regions of parameter space. 

\section{Extended DM Structures}
\label{sec:examples}
 
Extended DM configurations arise in many theories beyond the Standard Model and manifest in a variety of forms. 
We analyze Roman's sensitivity for three qualitatively distinct, well-motivated examples of extended DM distributions. Here, we consider each resulting extended lens configuration to be radially symmetric and characterized by its compactness, quantified in terms of $R_\text{90}$, the radius in which $90\%$ of the mass of the lens is enclosed. Though we choose to focus on three fiducial models, our analysis can be readily extended to account for other DM sources.

\subsection{NFW subhalos}

A hierarchy of DM halos ranging from galactic masses down to Earth-mass scales is a signature prediction of standard $\Lambda$-Cold Dark Matter ($\Lambda$CDM) cosmology. There is a large uncertainty on the low-mass end of this spectrum; hunting for these smaller subhalos provides a key test of the $\Lambda$CDM paradigm. 
Beyond this, various cosmological theories that extend the Standard Model predict  DM subhalo formation~\citep{Erickcek:2011us,Barenboim:2013gya,Fan:2014zua,Dror:2017gjq} at a range of scales. Characterizing the abundance of low-mass DM subhalos could therefore provide insight into physics beyond the Standard Model. 

Here, we choose to model subhalos via a characteristic Navarro-Frenk-White (NFW)~\citep{Navarro:1995iw} density profile that is known to fit simulations and galactic-scale observations
\begin{equation}
\label{eq:nfw}
\rho_\text{NFW}(r, \rho_s, \rs) = \frac{\rho_s}{(r/\rs)(1+(r/\rs))^2},
\end{equation}
where $\rho_s$ and $\rs$ denote the scale density and scale radius, respectively. While the total mass of this profile is formally divergent, we cut off the distribution at $100 ~\rs$ in keeping with existing literature~\citep{Croon_2020}. The compactness of the lens can be recast in terms of $\rn$, which is $\approx 69 ~\rs$ for this mass distribution.

While we choose to employ this distribution for our subhalos, our analysis can be readily extended to other profiles, such as those associated with secondary infall~\citep{Bertschinger:1985pd}. 

\subsection{Halo-dressed PBH-like configurations}

The unprecedented sensitivity of the GBTDS to point-like objects such as PBHs has been recently established 
in \citet{pruett_primordial_2022,DeRocco:2023gde}, and PBH microlensing in this mass-range can test a variety of intriguing theoretical models~(e.g.~\citet{Kusenko:2020pcg,DeLuca:2020agl,Sugiyama:2020roc,Kawana:2022olo,Lu:2022yuc,Inomata:2023zup}).
However, when PBHs constitute a subdominant fraction of DM, they can become surrounded by massive DM halos formed from the accretion of background DM via self-similar infall~\citep{Bertschinger:1985pd}; such a PBH is said to be ``dressed'' by its enclosing halo. The growth of this halo occurs primarily during matter-dominated era.

Numerical analyses \citep{1985ApJS...58...39B,Berezinsky_2013,Adamek_2019,Boudaud_2021,Serpico_2020} indicate that the resulting DM halos take a universal form well-described by the falling power law 
\begin{equation}
\label{eq:dressedpbh}
\rho_\text{dPBH}(r, \rho_s, \rs) = \rho_s\left(\frac{r}{\rs}\right)^{-9/4},
\end{equation}
where $\rho_s$ and $\rs$ are scale parameters. As with the NFW subhalo case, we can recast the scale radius $\rs$ into $\rn \approx 86.9 ~ \rs$. The mass of the halo $M_h$ and central black hole $M_{\rm BH}$ are related via \citep{Mack_2007,Ricotti_2008}

\begin{align}
\label{eq:dressedpbhrel}
M_h \simeq&~ 3\left(\frac{1000}{1+z_c}\right)M_{\rm BH} \notag\\
R_h \simeq&~ 0.019~{\rm pc}\Big(\dfrac{M_h}{1~M_{\odot}}\Big)^{1/3}\Big(\dfrac{1000}{1+z_c}\Big),
\end{align}

\noindent
where we take $z_c \simeq 30$ as the redshift at which the growth of the halo becomes inefficient due to interactions with nonlinear cosmic structures. 

We note that for masses of interest to the Roman GBTDS ($M_h \approx 10^{-9}-10\,M_\odot$), Eq. \ref{eq:dressedpbhrel} implies that $R_\text{90} \gg R_E$, where $R_E$ is the Einstein radius (see Sec.~\ref{sec:microlensing} below). Therefore, dressed PBHs with such compactness will ultimately be challenging to discriminate from bare PBHs by Roman. As stressed in \citet{Ricotti:2009bs}, microlensing lightcurves can efficiently distinguish dressed PBHs if the halo mass within an Einstein radius is at least a significant fraction of PBH mass, a criterion that is not met for these values of $R_{90}$. (We note that alternative lensing methods have been suggested to test dressed PBH scenario, especially around the solar mass range, including lensing of cosmological fast radio bursts~\citep{Oguri:2022fir} and gravitational waves~\citep{GilChoi:2023qrz}.) 

While this demonstrates that very diffuse DM objects will be difficult to probe with Roman, we adopt a similar strategy to \citet{Cai:2022kbp} and present results for PBHs enclosed by ``dressing'' for which the extent of the halo has been decoupled from the PBH mass for illustration. We therefore call this scenario ``dressed PBH-like,'' to indicate its phenomenological nature.

\subsection{Axion/boson stars}

Axions and, more generally, axion-like particles (ALPs), are pseudoscalar fields that are theoretically well-motivated (see e.g.~\citet{Adams:2022pbo} for review) and appear in many extensions of the Standard Model.
In the early universe, both ALPs and new scalar degrees of freedom have been shown to condense into macroscopic bound solitonic stars (see e.g.~\citet{Kaup:1968zz,Lee:1988av,Seidel:1993zk,Schunck:2003kk,Liebling:2012fv}). Among other formation scenarios, this is expected to occur generically within the cores of DM substructures such as diffuse axion miniclusters~\citep{Kolb:1993zz,Eggemeier:2019jsu}. Axion stars have been associated with a broad range of observational signatures, such as bosenova explosions~(e.g.~\citet{Eby:2016cnq,Levkov:2016rkk,Helfer:2016ljl,Eby:2021ece,Arakawa:2023gyq}) as well as microlensing \citep{Fujikura_2021}. 

The density profile of an axion or boson star does not afford a closed-form analytic solution and must instead be calculated numerically by solving the coupled Schrodinger-Poisson equations. Rather than solving this for a particular microphysical model, we instead adopt a phenomenological model of the the ground state solution, which has been shown to be well-described by \citep{Schiappacasse_2018}
\begin{equation}
\label{eq:axionstar}
\rho_\text{ax}(r, \rho_s, \rs) = \frac{3 \rho_s}{\pi^3 \rs^3} \text{sech}^2\left(\frac{r}{\rs}\right).
\end{equation}
For this profile, we find $\rn \simeq 2.8 ~\rs$ and truncate the profile at $20 ~\rs$. 

In contrast to the dressed PBH scenario, simple axion models provide a mass-radius relation for stable axion stars \citep{Schiappacasse_2018,Visinelli_2018,Sugiyama_2023} that lead to $R_{90} \ll R_E$. Hence, this particular subclass of axion/boson star appears effectively point-like to Roman and the GBTDS will be particularly sensitive to these scenarios.

\section{Microlensing}
\label{sec:microlensing}

Gravitational microlensing \citep{paczynski_gravitational_1986}, the apparent magnification of a luminous source by the gravitational field of a lensing mass, is one of the strongest observational probes of macroscopic dark matter in the mass range $\approx 10^{-11} \,M_\odot - 10\,M_\odot$~\citep{Niikura:2017zjd,Niikura:2019kqi,MACHO:1998qtf}. 
This magnification depends on the impact parameter of the event $u$, the transverse distance \textbf{in the lensing plane} between the center of the source and center of the lens. Throughout this work, we normalize the impact parameter by the Einstein radius of the lens,
\begin{equation}
\label{eq:RE}
R_E = \sqrt{\frac{4 G M D_L (1 - D_L/D_S)}{c^2}},
\end{equation}
where $M$ is the mass of the lens and $D_S$ and $D_L$ are the distance from the Earth to source and lens, respectively. This quantity sets the typical transverse scale in the plane of the lens over which the lensing effect is appreciable, $\theta_E = R_E/D_L$. The angular size of the source is likewise given by $\theta_\star = R_\star/D_S$, where $R_\star$ is the radius of the source.

When both the source and the lens can be approximated as points (the ``point-source point-lens'' (PSPL) regime), the magnification curve follows a well-known analytic form given by
\citep{nakamura_wave_1999} 
\begin{equation}
    A_\text{PSPL}(u) = \frac{u^2 + 2}{u\sqrt{u^2 + 4}},
    \label{Ageo}
\end{equation}
where $A$ is the observed magnification.
In general, however, the angular size of the source and lens may not be negligible. The magnification can no longer be computed analytically and one must instead compute the point-source, finite-lens magnification $A_\text{PSFL}$ numerically, then integrate this over the angular extent of the source. One can derive $A_\text{PSFL}(u)$ via the \textit{lensing equation}:
\begin{equation}
    \label{eq:lensingeq}
    u = v - \frac{1}{v}\frac{M(v)}{M_{tot}}
\end{equation}
where $v$ is the transverse distance from the center of the lens to the image, once again normalized to $R_E$, and $M(v)$ is the projected mass contained within a radius $v$ in the plane of the lens, hence
\begin{equation}
    \label{eq:mprojected}
    \frac{M(v)}{M_\text{tot}} = \frac{\int_0^v dv'~ v' \int_{-\infty}^\infty dz~\rho(\sqrt{z^2 + v'^2})}{\int_0^\infty dv'~ v' \int_{-\infty}^\infty dz~\rho(\sqrt{z^2 + v'^2})}.
\end{equation}
Here, $\rho(r)$ is the radial mass distribution of the lens. As discussed in the previous section, this function may take many forms, but is typically controlled by a scale parameter that determines the lens compactness. Here, we choose $\rne$, the radius in which $90\%$ of the mass is enclosed, in units of the Einstein radius. Note that we adopt the convention of lower-case names for length scales in units of $R_E$ or $\theta_E$, e.g. $\rn = \rne R_E$.

\begin{figure*}
    \centering
    \includegraphics[width=0.65\linewidth]{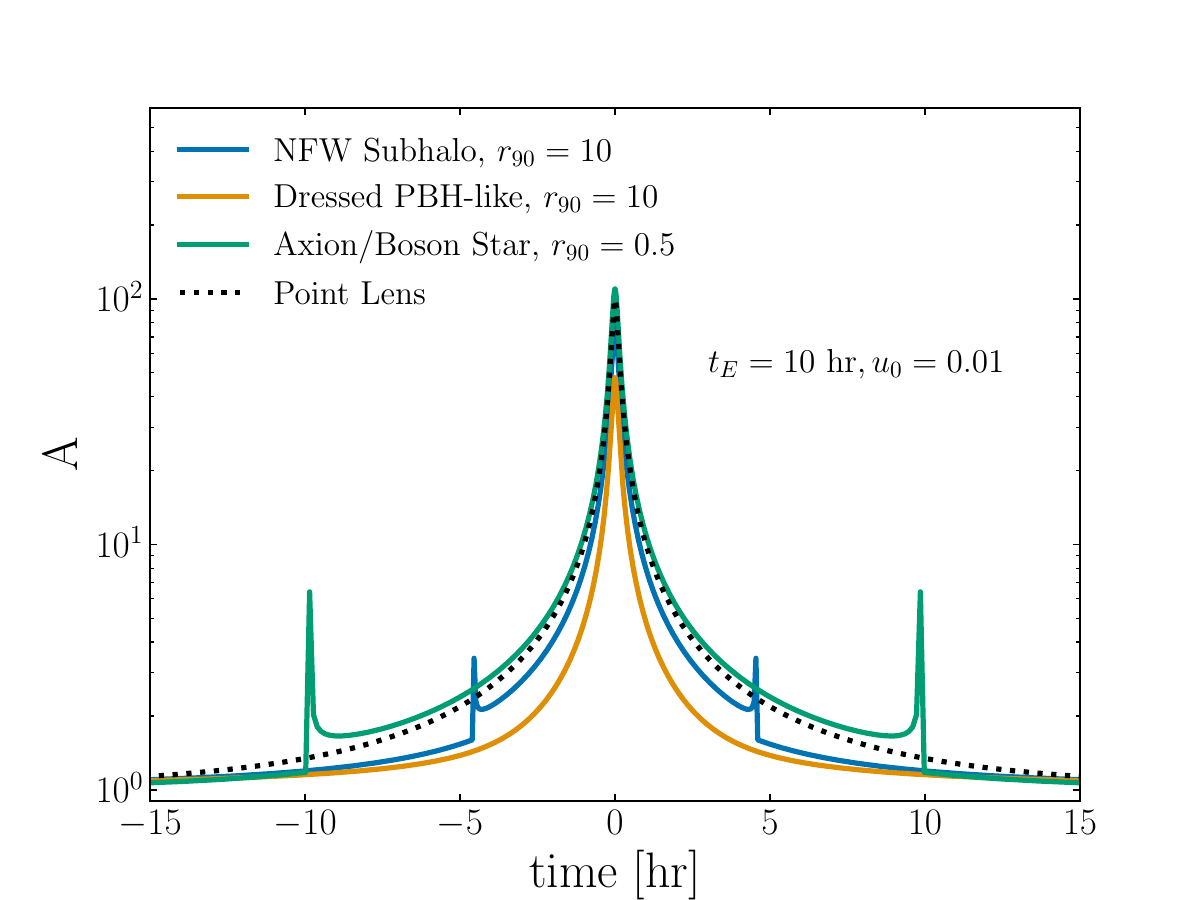}
    \caption{Representative light-curves for the three extended lens species (colored lines) plotted alongside a point lens (dotted line), all in the point-source regime. For certain parameter choices, the extended lenses can produce caustic features, seen here in the case of the NFW subhalo (blue) and axion/boson star (green). See discussion below Eq. \ref{eq:Apsfl}.}
    \label{fig:caustics}
\end{figure*}

For a given mass distribution, Eq.~\ref{eq:lensingeq} can be solved to yield $v(u)$, the image location as a function of the impact parameter. This function is often multivalued, with different branches corresponding to different images. Therefore, one must compute the magnification associated with each image separately and sum them to find the total magnification. This yields
\begin{equation}
    \label{eq:Apsfl}
    A_{\text{PSFL}}(u, r_{90}) = \sum_i \left|\frac{v_i(u)}{u}\frac{dv_i(u)}{du}\right|,
\end{equation}
where $i$ denotes a particular image. At points where the slope of $v(u)$ becomes infinite, the magnification diverges. This feature is known as a \textit{caustic} (see e.g. \citet{Karamazov:2021hwa, Hurtado:2013zda} for discussion of their presence in the light-curves of spherical extended lenses). 
If these caustic features can be resolved, their location and shape contain information about the underlying mass profile, providing a unique opportunity to uncover the microphysical nature of the lens. We illustrate this possibility in Fig. \ref{fig:caustics}. However, we leave a full characterization of Roman's sensitivity to extracting these features to future work.

Having numerically calculated $A_{\text{PSFL}}$, the ultimate magnification curve, taking into account both finite-source effects and the extended lens, is given by an integral over the angular extent of the source. Working in polar coordinates $(r, \phi)$ with the origin aligned to the center of the source, this is given by
\citep{matsunaga_finite_2006, witt_can_1994, sugiyama_revisiting_2019}
\begin{equation}
    A_{\mathrm{FSFL}}(u,\rho, \rne) \equiv 
    \frac{1}{\pi \rho^2} \int_0^{\rho} dr \int_0^{2\pi} d\phi ~r 
    ~A_{\rm{PSFL}}\Big( \sqrt{r^2 + u^2 - 2ur \cos(\phi)}, \rne \Big).
    \label{Afinite}
\end{equation}
where $\rho \equiv \theta_\star/\theta_E$.

Given this curve, one can implicitly solve for the maximum impact parameter $u_T$ that yields a ``detectable'' event, where we define detectability as exceeding some threshold magnification $A_T \equiv A_\text{FSFL}(u_T)$. This sets the integration range over which we compute our event rate, as discussed in the following section.

\section{Event rate}
\label{sec:analysis}

Here, we describe our methodology for estimating the expected number of extended-lens events that Roman will detect during the GBTDS. 
We begin by computing the differential expected rate of detectable events, which is given by \citep{batista_moa-2009-blg-387lb_2011}
\begin{equation}
    \label{eq:difRateAll}
     \frac{d\Gamma}{dD_L\,dt_\text{dur}\,du_0} =  
     f_{\rm{DM}} \frac{2}{\sqrt{u_T^2 - u_0^2}} \frac{v_T^4}{v_c^2} \exp \Big[ -\frac{v_T^2}{v_c^2}\Big] \frac{\rho_M}{M} \varepsilon(t_\text{dur}),
\end{equation}
where $f_{\rm{DM}}$ is the fraction of DM mass density composed of the extended-lens population of interest, $\rho_M$ is the mass density of lenses, $u_0$ is the impact parameter at the point of closest approach, and $\varepsilon(t_\text{dur})$ is the detection efficiency \citep{Johnson_2020}. Here $v_c$ is the typical velocity dispersion of the dark lenses, calculated assuming the distribution of dark matter in the Galactic halo follows an NFW profile (see Eq. \ref{eq:nfw})
\begin{equation}
    v_c(r) = \sqrt{\frac{G M_{\rm{NFW}}(<r)}{r}}.
\end{equation}
$M_{\rm{NFW}}(<r)$ is the mass enclosed within radius $r$ of the Galactic center, where we take $\rho_0 = 4.88 \times 10^6 ~M_{\odot} ~\text{kpc}^{-3}$ and $\rs = 21.5 ~\text{kpc}$ following \citet{klypin_lcdm-based_2001}.
The relative source-lens velocity is given by 
\begin{equation}
\label{eq:vT}
 v_T = 2\theta_E D_L \sqrt{u_T^2 - u_0^2}/t_\text{dur},
\end{equation}
where the duration of the lensing event, $t_\text{dur}$, is defined as the longest continuous interval for which the magnification exceeds the threshold for detection ($A>A_T$). Often $A_T = 1.34$ is chosen as a fiducial threshold as it corresponds to $u_T = r_E$ in the PSPL limit. However, this threshold is likely overly conservative for Roman, which is expected to achieve photometric sensitivity sufficient to resolve percent-level changes in flux \citep{Johnson_2020}. 

As such, we determine $A_T$ by using a more realistic detection criteria that is consistent with existing analyses of Roman sensitivity \citep{Johnson_2020}. We require at least $6$ measurements with magnifications at least $3\sigma$ above the baseline, where the baseline depends on both the source flux and blending from neighboring stars. This corresponds to 
\begin{equation}
    A_T = \frac{1 + 3 \sigma - \Gamma_c}{1 - \Gamma_c},
\end{equation}
where $\Gamma_c \equiv \frac{f_b}{f_s+ f_b}$ is the \textit{flux contamination}, indicating the fraction of the total flux contributed by the source's neighbors, and $\sigma$ is the magnitude-dependent photometric precision, which we take from Figures 5 and 9 of \cite{wilson_transiting_2023}, respectively. Finally, we weight the magnification-dependent threshold by the stellar population of the GBTDS using Figure 15 of \cite{wilson_transiting_2023}, resulting in a population-weighted detection threshold of $A_T \approx 1.05$.

The total detected event rate, $\Gamma$, is then calculated by integrating 
\begin{equation}
    \label{eq:rate}
     \Gamma= 2 \int_0^{D_S}dD_L \int_0^{u_T}du_0 \int_{t_\text{min}}^{t_\text{max}}dt_\text{dur}
     f_{\rm{DM}} \frac{1}{\sqrt{u_T^2 - u_0}^2} \frac{v_T^4}{v_c^2} \exp \Big[ -\frac{v_T^2}{v_c^2}\Big] \frac{\rho_M}{M} \varepsilon(t_\text{dur}),
\end{equation}
where $t_\text{min} = 90 ~\rm{min}$ and $t_\text{max} = 72 ~\rm{days}$. The lower cutoff, $t_\text{min}$, corresponds to $6 \times 15\,\text{min}$, which is necessary to satisfy our detection criterion of 6 measurements with $A > 3 \sigma$, where 15 minutes is the nominal cadence of the GBTDS. The higher cutoff, $t_\text{max}$, is the proposed observation season duration for the GBTDS. This has still yet to be determined, though it is constrained to be between 60 and 72 days \citep{wilson_transiting_2023}. As such, we adopt the current nominal season duration of 72 days and display how shortening the season would influence our results with the dashed curve in Fig. \ref{fig:results_all}. Regardless of choice for $t_\text{max}$, the total expected number of events is then simply given by $N = 6\times\Gamma t_\text{max} $, due to the 6 observation seasons.

If the astrophysical backgrounds are fully known, i.e. no detected events can be attributed to an extended lens, then we determine the maximum sensitivity of Roman to a given population under the condition of null detection. In practice, it is often not possible to determine the identity of a lens on an event-by-event basis. Therefore, our results are not a detailed projection of expected limits, but rather a demonstration of Roman's maximum sensitivity to extended lenses.
The Poisson 95$\%$ confidence limit under the assumption of null detection corresponds to $N \leq 3$, which yields a maximum allowed value for $f_{\rm{DM}}$. We perform the calculation of the event rate using the open-source Python package \texttt{LensCalcPy} \citep{smyth_2024_10734644}, a flexible tool for estimating microlensing rates that has been shown to yield results consistent with full population synthesis models \citep{DeRocco:2023gde}.

\section{Results}
\label{sec:results}

\begin{figure*}
    \centering
    \includegraphics[width=\linewidth]{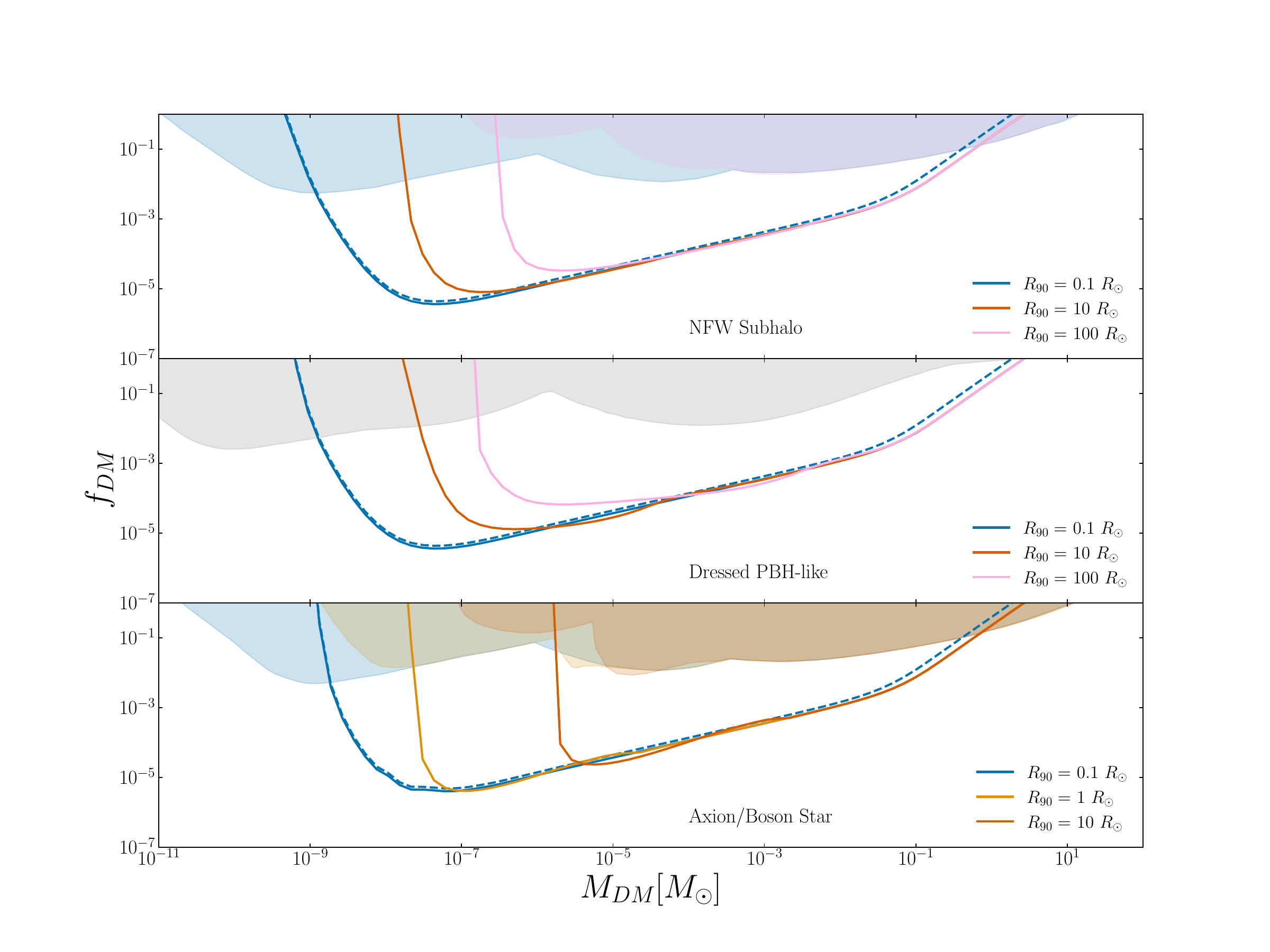}
    \caption{Projected sensitivity to different extended lenses. Solid lines are estimates for 72-day seasons. The dashed lines show the projected sensitivity to $R_{90} = 0.1 ~R_{\odot}$ lenses using 60-day seasons. Existing limits from Subaru, OGLE, and EROS are shown in shaded regions \citep{croon_subaru_2020} with the color corresponding to the extent of the lens. The gray shading corresponds to limits on point-like lenses with mass $M_\text{dPBH} \approx 100 ~M_\text{PBH}$, corresponding to the most optimistic current constraints on dressed PBH (see \citet{Cai:2022kbp} for details).}
    \label{fig:results_all}
\end{figure*}

Our results are displayed in Fig. \ref{fig:results_all}, where the horizontal axis shows the total mass of the lens and the vertical axis shows the mass fraction of dark matter contributed by a population of lenses at that mass. The colors of the curves correspond to various values of $\rn$ ranging from $\rn = 0.1 ~R_{\odot}$ (effectively a point-like lens) up to $\rn = 100~ R_{\odot}$ (a highly-extended lens). The filled-in regions correspond to existing constraints from other microlensing surveys \citep{croon_subaru_2020,Cai:2022kbp}. 

We find that Roman will be most sensitive to lens masses of $\sim 10^{-7} \, M_\odot$, corresponding to event durations on the timescale of hours. In general, for more diffuse lenses (larger $\rn$), the sensitivity weakens. This is particularly noticeable at low masses in which the magnification peak departs from the quasi-singular PSPL curve (Eq. \ref{Ageo}). At high masses, the curves converge since the lenses are effectively point-like, even for lenses with $\rn = 100 ~R_{\odot}$. At $M \gtrsim 10^{-3}\, M_\odot$, the various curves all begin to scale as $M^{-1/2}$. In this regime, the event rate is linearly proportional to the number density of lenses and the cross-section for a detectable event; since the number density for a monochromatic population scales as $1/M$, while the Einstein radius, which effectively defines the cross-section of microlensing events in the point-source limit, scales as $\sim M^{1/2}$, the resulting rate dependence scales as $M^{-1/2}$.

The sharp cutoff at low masses corresponds to the point at which finite-source effects suppress the peak magnification sufficiently to render events undetectable. The maximum magnification in this regime scales as $A_{\rm{max}} = \sqrt{1 + 4/\rho^2}, ~(\rho \gtrsim 1)$. The cutoff in mass corresponding to this condition depends on both the compactness of the lens, which changes the effective $R_E$ for a given total mass, and the sensitivity of the instrument, which determines $A_T$. For a point-like lens, this cutoff translates to 
\begin{equation}
    M_{\rm{cut}} \approx \frac{(A_T^2 - 1) D_L R_\star^2 c^2}{16 D_S^2 G},
\end{equation}
where it is assumed $D_L \ll D_S$. This corresponds to $M_{\rm{cut}} \approx 8.7 \times 10^{-10} M_{\odot}$ for $A_T = 1.05$, $D_L = 1 ~\rm{kpc}$, $D_S = 8.5 ~\rm{kpc}$, and $R_\star = ~R_{\odot}$, which coincides well with the corresponding cutoff in the figure.

The dashed curve shows the sensitivity for $R_{90} = 0.1 R_\odot$ if the Roman observational season duration is shortened from 72 to 60 days (see Sec. \ref{sec:analysis}). Since the peak sensitivity to extended lenses lies at $\sim$hour-long event durations, this change does not have an appreciable affect on our results, with sensitivity weakening by a factor of $(72/60) \approx 1.2$ at low masses and $(72/60)^4 \approx 2$ at high masses due to the reduced phase space.

\section{Conclusions}
\label{sec:disc}

A wide variety of motivated theoretical models predict the formation of extended substructures in the dark sector. For the first time, we have comprehensively analyzed the sensitivity of the upcoming Roman Space Telescope to detecting such substructures in the Galactic Bulge Time Domain Survey. Our results, illustrated for three well-motivated but qualitatively distinct extended DM targets, show that Roman will be able to explore a broad parameter space at up to four orders of magnitude lower fractional DM mass contributions than existing constraints. Furthermore, the methodology outlined in this paper can be readily extended to other sources.
We note that the presence of caustic features in light-curves will provide future opportunities to discriminate extended substructures from point-like DM candidates, providing Roman with the potential to not only detect DM substructures, but to probe their microphysical nature as well. We leave further exploration of this exciting possibility to future work. Furthermore, our work strongly motivates the GBTDS to adopt the maximal season duration subject to mission constraints (72 days); this result has immediate relevance for other targets in this mass range such as free-floating planets and isolated black holes as well.

\begin{acknowledgments} \textit{Acknowledgments---} WD and NS acknowledge the support of DOE grant No. DE-SC0010107. VT acknowledges the support by the World Premier International
Research Center Initiative (WPI), MEXT, Japan and the JSPS KAKENHI grant 23K13109.
\end{acknowledgments}

\bibliography{ref}

\end{document}